\documentclass{aa}             % <-- "class" anstelle von "style"
\usepackage{amssymb,graphicx}  % <-- "graphicx" fuer Bilder, "amssymb"
\begin{document}

\title{The effect of disk magnetic fields on the truncation of
geometrically thin disks in AGN}
\author{F. Meyer  and E. Meyer-Hofmeister }
\offprints{Emmi Meyer-Hofmeister}
\institute{Max-Planck-Institut f\"ur Astrophysik, Karl-
Schwarzschildstr.~1, D-85740 Garching, Germany
} 

\date{Received: / Accepted:}

\abstract{
We suggest that magnetic fields in the accretion disks of AGN reach
into the coronae above and have a profound effect on the
mass flow rate in the corona. This strongly affects the location
where the accretion flow changes from a geometrically thin disk to a
pure vertically extended coronal or advection-dominated accretion
flow (ADAF). We show that this can explain the different disk
truncation radii in elliptical galaxies and low luminosity AGN with about
the same mass flow rate, a discrepancy pointed out by Quataert et
al. (1999).
Without disk magnetic activity the disk truncation is expected to
be uniquely related to the mass flow rate (Meyer et al. 2000b).
Whether dynamo action occurs depends on whether the electrical
conductivity measured by a magnetic Reynolds number surpasses a
critical value (Gammie \& Menou 1998).
In elliptical galaxies the disk is self-gravitating at the radii where
the truncation should occur. It is plausible that instead of a cool disk
a ``layer of clouds'' may form (Shlosman et al. 1990, Gammie 2001) for
which no dynamo action is
expected. For low luminosity AGN the magnetic Reynolds number is well
above critical. Simple
model calculations show that magnetic fields in the underlying disks 
reduce the strength of the coronal flow and shift the truncation
radius significantly inward.
}
\titlerunning {Magnetic fields and thin disk truncation in AGN}
\maketitle
%
%  1 figure: Ee163_f1.ps
%_____________________________________

\section{Introduction}
Accretion onto supermassive black holes can occur via an
advection-dominated accretion flow (ADAF) or via a standard 
geometrically thin accretion disk. The mode depends on the accretion
rate. In an ADAF advection rather than radiation removes the locally
generated accretion heat.
This is only possible in the inner region around the central accretor.
At larger distance the mass accretes via a cool geometrically thin
standard disk. The situation is the same for supermassive and
galactic black holes (for a review on the now already ``classical''
advection-dominated accretion and applications see Narayan
et al. 1998). Confirmation for this picture, an ADAF in the inner
region near the black hole and a surrounding cool standard disk with a
corona above came from the successful modeling of the spectral transitions of
low-mass X-ray binaries (LMXRBs) which arise from the changing mode of
accretion (Esin et al. 1997, 1998).

The change from disk accretion to an ADAF was investigated for 
several low luminosity AGN. The low luminosity in the elliptical
galaxy M87 (NGC 4486) was pointed out by Reynolds et al. (1996).
Quataert et al. (1999) investigated the low-luminosity galactic nuclei
of M81 and NGC 4579 and from the spectral fits found evidence for thin
disks truncated at about 100 Schwarzschild radii, similar to the
result for NGC 4258 (Gammie et al. 1999). Di Matteo et al. (1999,
2000) studied the nuclear regions of six elliptical galaxies and
derived larger truncation radii from the X-ray and
high-frequency radio observations. In these fits wind loss was
included which seems to be demanded at least in some
galaxies. This means a significant fraction of mass, angular momentum
and energy is removed from the accretion flow by a wind. Large radii
were found also for M 87 and NGC 4696 by Quataert et al. (1999)
(see detailed discussion of constraints for the different objects in their
paper). That mass loss via wind may be important was also stated
by Quataert \& Narayan (1999), who pointed that different combinations
of parameters characterizing wind loss and micro physics lead to
equally good models. That means comparably good fits to
the spectra can be found for different mass flow rates and truncation
radii. But despite this ambiguity the fits showed, that there is a
clear discrepancy of truncation radii in nuclei of elliptical galaxies
of order $10^3$ to $10^4 r_{\rm {S}}$ and those in low
luminosity AGN (LLAGN) of only $10^2 r_{\rm {S}}$ as pointed out by
Quataert et al. (1999) (truncation radii are measured in Schwarzschild
radius $r_{\rm S}=2GM/c^2$, $M$ mass of central black hole).

Since the transition to the ADAF can be understood as caused by
evaporation of mass from the thin cool disk to a hot
corona above, the truncation of the thin disk depends on the
amount of mass flow there (Meyer \&
Meyer-Hofmeister 1994, Meyer et al. 2000b). If one measures the
radii $r$ in Schwarzschild radius and the mass accretion rate
$\dot M$ in Eddington
mass accretion rates ($\dot M_{\rm {Edd}}=40{\pi}GM/{\kappa_{\rm {es}}}c$,
$\kappa_{\rm {es}}$ electron scattering opacity) theory predicts a unique
relation between truncation radius and mass accretion rate.
But in fact the disks in LLAGN reach much closer in, to only 100
Schwarzschild radii, demanded by the observed UV radiation while the
accretion rates ${\dot M/\dot M_{\rm {Edd}}}$ are comparable to those
in the nuclei of elliptical galaxies.

In the following we suggest that the presence or absence of a disk
dynamo explains the observed difference of truncation radii between
LLAGN and elliptical galaxies with the same mass accretion rate.
Sect. 2 gives a short description of the transition from disk
accretion to a coronal flow. In Sect. 3 we investigate the presence or
absence of disk dynamos in AGN. In Sect. 4 we present results of
our computations including a magnetic field. A discussion follows in
Sect. 5, conclusions in Sect. 6.

\section {The transition from thin disk accretion to an ADAF caused by
evaporation}

In previous work (Meyer et al. 2000b) we presented a model for the
corona above a geometrically thin standard disk around a black hole.
This corona is fed by matter continuously evaporating from the cool
disk. The accretion flow is thus divided into a hot coronal flow and
a part remaining in the cool disk. The strength of the coronal flow
increases inward so that at a certain distance in the inner region
all matter is transferred to the coronal flow and the disk is truncated.
The situation is the same for galactic and supermassive black holes.
The theoretically derived relation for the dependence of evaporation
rate on black hole mass and distance allows to determine the
truncation radius for each accretion rate. This relation
was successfully applied to X-ray transients. Liu
\& Meyer-Hofmeister (2001) discussed the application to AGN
and found reasonable agreement, except that the
truncation radii for LLAGN clearly were much smaller than predicted.

\section {Disk dynamos in AGN}
Up to now the evaluation of the transition from disk accretion to a 
coronal flow/ADAF has neglected a possibly important aspect:
If the temperature in the underlying cool disk is high enough 
for dynamo action to occur the magnetic fields generated will 
penetrate also into the corona and affect the coronal accretion
flow. Magnetic dynamos require sufficiently long Ohmic decay times compared
to the dynamical time of the dynamo process. This ratio is measured by
a magnetic Reynolds number, which is strongly temperature dependent for low
temperatures.

 Numerical simulations by Hawley et al. (1996) determined
a critical value of the magnetic Reynolds number of about $10^{3.5}$.
Below this
value dynamo action becomes significantly suppressed. Gammie \& Menou (1998)
showed that dwarf nova accretion disks in quiescence are sufficiently
cool to be below this critical value. AGN accretion disks have much
larger physical dimensions. Menou and Quataert (2001) have
demonstrated that the magnetic Reynolds numbers in such disks are above
the critical value in the corresponding quiescent
state. Here we are concerned with even further out ranges in
elliptical galaxies where temperatures become extremely low. 

In elliptical galaxies the black hole masses are about $10^9 M_\odot$
or more, in LLAGN only a few $10^6 M_\odot$. For
higher black hole
masses the midplane temperature is lower when normalized radii
$r/r_{\rm {S}}$ and accretion rates $\dot M/\dot M_{\rm {Edd}}$ are
kept the same.  As an example we consider the situation for a mass flow rate
$10^{-2.5} \dot M_{\rm{Edd}}$ as derived for M81.
For this rate we consider the disk structure at a radius of
$10^{3.5}r_{\rm {S}}$ where the truncation would be expected
from the standard curve (Fig.1).

Structure and evolution of accretion disks in AGN were investigated
mainly in connection with the ionization instability in
these disks (Mineshige \& Shields 1990, Cannizzo \& Reiff 1992,
Cannizzo 1992, Siemiginowska et al. 1996). The radii of interest for
our evaluation are larger than those considered there. For high
central masses self-gravitation becomes important at such radii.

To explore a possible influence of disk dynamo action on the corona we
refer to investigations for self-gravitating disks. Shlosman et
al. (1990) discussed an attractive model of a ``disk of clouds'':
when the angular momentum transport is locally mediated gravitational
interaction, the disk breaks up and possibly forms a layer of clouds
(which should occur if the cooling time is short compared to the Kepler time).
Recent numerical computations of Gammie (2001) for a thin horizontally
extended layer confirm this picture with the formation of blobs.
The clouds then by interaction have to transport the angular momentum, no
effective dynamo would be expected. How can in this case the accretion
flow via the clouds be transformed to a coronal flow ?
According to the standard evaporation model these clouds must be
embedded in a corona in equilibrium with the cool surfaces of the blobs.
Thermal conduction of the hot corona to the cool surfaces establishes
an equilibrium density and a mass flow rate from the blobs to the
coronal gas, similar to the case of an underlying cool
disk. The conductive flux requires a large enough surface area.
Estimates give that a cloud covering factor $C\geq 10^{-2}$ is required,
which seems
possible. Then the cloud layer is cut off and closer to the center
only a coronal flow (ADAF) exists.

Balbus \& Papaloizou (1999) discuss an alternative picture, in which
waves propagating over long-range distances transport angular
momentum and also energy (note that Gammie (2001) gives arguments
against the formation of large scale coherence). The effective removal
of angular momentum in
this way might allow the existence of a disk marginally stable
against break up. Because of non-local energy transport the disk
would be cooler than a corresponding ``high-$\alpha$'' disk with the
same angular momentum transport. For the latter disk (for a marginally
stabilized disk see Paczy\'nski 1978) in our case the formal values
of  $\alpha$ would be $\geq 10$. The midplane temperature in such a disk
can be derived using a technique for solving the full vertically
averaged disk equations similar to the accretion disk modeling in
Cannizzo \& Reiff (1992). We get for the temperature
\begin{equation}
T=10^{2.7} \big(\frac{M}{10^9 M_\odot})^{-4/7}
           \big(\frac{r}{10^{3.5}r_{\rm {S}}})^{-9/7}
           \big(\frac{\dot M}{10^{-2.5} \dot M_{\rm{Edd}}}) \rm{K}
\end{equation}
and for the density
\begin{equation}
\rho = 10^{-12.6}\big(\frac{M}{10^9 M_\odot})^{-2}
                 \big(\frac{r}{10^{3.5}r_{\rm {S}}})^{-3} \rm{gcm}^{-3}.
\end{equation}

The value of $\alpha$ is then dependent on $M$, $r$ and $\dot
M$, incorporated in the formulae above. In the derivation of Eqs. (1)
and (2) we used dust opacities, $\kappa=1\, cm^2/g$. The heating in
the disk is due to internal dissipation.Direct and indirect (by
scattering on the clouds) irradiation is negligible for the example M 87
because of the low luminosity.

With these numbers we can evaluate the magnetic Reynolds number 
\begin{equation}
Re_{\rm{M}} \equiv V_{\rm{s}}H /\eta =
\sqrt{2}\frac{\Re}{\mu} T/ \Omega^2\eta
\end{equation}
with $V_{\rm{s}}$ isothermal sound velocity, $H$ disk scale height,
$\eta$ magnetic diffusivity, $\Re$ gas constant, $\mu$  molecular weight,
$\Omega$ Kepler rotation frequency.

With the magnetic diffusivity $\eta =m_e c^2 \nu_{en}/n_e e^2$
($n_e$ electron number density, $e$ electron electric charge,
$m_e$ electron mass, $\nu_{en}$ electron neutral collision frequency 
proportional to the number of neutral particles $n_n$) the Reynolds
number becomes proportional to the ratio $x_e=n_e/n_n$ (Gammie \&
Menou 1998). This gives
\begin{equation}
Re_{\rm{M}}=10^{9.7} \big(\frac{M}{10^9 M_\odot})
            \big(\frac{r}{10^{3.5}r_{\rm {S}}})^3 [X(T)\cdot T]^{1/2} 
\end{equation}
with $X(T)=n_n x_e^2$ is a function of temperature only determined by
the Saha equation. The ionization state of the electron providing alkali
metals becomes extremely low for the low temperatures.
This yields a Reynolds number below the critical value of $10^{3.5}$
for $T \le 10^{2.9}$ K. At the low densities of these disks ambipolar
diffusion (Cowling 1976) increases the magnetic diffusivity (see also
recent simulations by Sano \& Stone 2002) and
thereby reduces the magnetic Reynolds number.
With typical dynamo values $\beta \approx 20$ ($\beta$ ratio of gas
pressure to magnetic pressure) the reduced  
$Re_{\rm{M}}$ is below the critical value for $T \le 10^{3.1}$ K.
Then no effect of disk magnetic fields is expected.

We conclude that in both these cases, a disk of clouds and a disk
marginally stabilized by wave transport, the evaporation of cool gas
into the corona occurs without the influence of a disk magnetic field.

For the disks in LLAGN we derived the temperature in the same
way (such disks are not self-gravitating and we took $\alpha$=0.01).
At the same (scaled) distance and mass accretion rate one obtains
$T=10^{3.3}$ K. This is in the range already considered
by Menou \& Quataert (2001). In agreement with their analysis
the magnetic Reynolds number becomes large enough to allow dynamo
action.

\section {The effect of the disk magnetic field on the truncation
radius}
We investigated how a magnetic field in the thin disk affects the coronal
mass flow. The amount of matter
that has evaporated from the disk and flows in the corona at a given
distance from the black hole can be determined by solving a set of ordinary
differential equations which describe vertical dynamical equilibrium,
conservation of mass and energy and heat flux (Meyer et al. 2000b). 
Our earlier computations take magnetic fields of a
coronal dynamo only implicitly, as a source of friction, into account
as in standard disk theory. But an analytical estimate was given how
$\beta$ enters into numerical results (Meyer et al. 2000b).
For the present investigation we explicitly model the effect of
additional magnetic fields (e.g. from an underlying disk) on coronal
dynamical equilibrium, energy release, and diffusive inward mass flow.
To a first approximation all effects can be
accounted for if in all equations one formally replaces the gas
pressure $P$ by the product $P (1+1/\beta)$.
The result is a remarkable reduction of the strength of
the coronal mass flow, i.e. of the evaporation efficiency, with
increasing magnetic field.

\begin{figure}[ht]
\includegraphics[width=8.8cm]{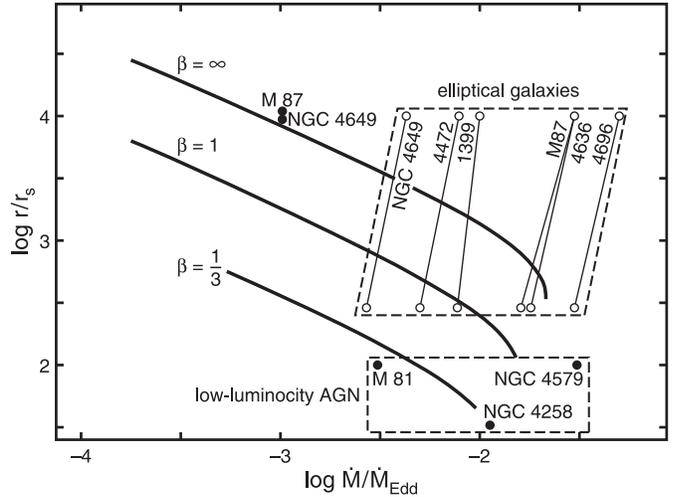}
\caption {The dependence of the truncation radius on the magnetic field
strength.
Solid lines: theoretical $r$-$\dot M$ relations for different ratios
$\beta$ = gas pressure/magnetic pressure.
%Theoretical $r$-$\dot M$ relations for different ratios
%$\beta$ = gas pressure/magnetic pressure (black solid line:
%without magnetic field, red lines: with magnetic field).
- For comparison results for $\dot M$ and truncation radius from fits to
observed spectra.
Open circles (fits including wind loss): 
elliptical galaxies from Di Matteo et al. (2000) (best and second-best
fit); filled circles (without wind loss): NGC 4649 from Quataert \& Narayan
(1999), M87 from Reynolds et al. (1996), same value also from
Di Matteo et al. (2000), LLAGN M81 and NGC 4579 from
Quataert et al. (1999), NGC 4258 from Gammie et al. (1999).
}
\end{figure}

Matter therefore remains in the cool disk down to much smaller radii
before the coronal flow has picked up all the accretion flow
and the disk is truncated. Fig. 1 shows the truncation radius
as a function of the distance from the black hole. The values are
scaled to $\dot M_{\rm {Edd}}$ and  $r_{\rm {S}}$.
(Inside of the truncation radius all matter flows in the form of
a hot corona). The additional field shifts the
truncation radius to smaller distances $r$.
A moderate magnetic field strength in the corona
($\beta$=1)
reduces the truncation radius by about a factor 5, a three times
higher magnetic pressure results in a reduction by a factor 25.
This is an important change.

In addition Fig. 1 shows data derived from observations by various
authors. The observational data show a wide range of truncation
radii. In particular only radii of $10^4$ or 300 $r_{\rm S}$ were
considered in the investigation of nuclei of elliptical galaxies (Di
Matteo et al. 2000) dealing with spectral fits based on
an extension of the ADAF model including loss of mass and angular
momentum by a wind. Note that spectral fits with wind loss yield only
lower limits of truncation radii. For M 87 both, models with and without
wind loss agree with the observed spectra. (Different results for NGC
4649 are due to different black hole mass.) 
Despite this spread in acceptable $r$ - $\dot M$ combinations
there is a clear discrepancy between the truncation radii for LLAGN and
nuclei of ellipticals with comparable mass flow rates (Quataert et al. 1999).

\section {Discussion}
\subsection{Dynamos in AGN disks}

We expect dynamo action for LLAGN, but not for elliptical galaxies
with a very high central mass. Otherwise if accretion rates are low
it is possible that the temperature even in disks around $10^6
M_\odot$ is low enough to forbid dynamo action (an example could be Sgr A*
with a central mass of $2.5\cdot 10^6M_\odot$ indicated by
observations (Genzel et al. 1997) and an accretion rate from
spectral fits based on ADAF models of 
$10^{-4} \dot M_{\rm{Edd}}$ (Quataert \& Narayan 1999)  - if
the thin disk really exists, doubts come from new Chandra
observations (Narayan 2002)).
On the other hand for higher mass accretion rates disks around the high
mass black holes in elliptical galaxies are hot enough to allow dynamo
action at the standard truncation radius, so that the true truncation
radius becomes shifted inward to smaller radii.

\subsection{The strength of the disk magnetic fields in the corona}
To estimate the field strength of disk dynamo fields in the corona is
difficult. The data for M 81 (Fig. 1) indicate a value of $\beta \approx
1/3$ as appropriate. Near the truncation radius mass flow rates in
corona and disk are of comparable size. This implies that the product 
$(1+\frac{1}{\beta})HP$ is about the same. From this one can estimate
the required magnetic pressure in the corona $(\frac{B^2}{8\pi})_{\rm
{corona}} \approx  10^{-1.2}(\frac{B^2}{8\pi})_{\rm {disk}}$ where the
temperature ratio $T_{\rm{disk}}/T_{\rm{corona}}\approx 10^{-4.6}$
estimated for M 81
and $(\beta)_{\rm{disk}}=20$ were used. Such values may be reached if
a corresponding fraction of the dynamo magnetic energy is cascaded to
scales of the coronal scale height or larger (e.g. Arlt \& Brandenburg 2001).

\section {Conclusion}

The investigation of the effect of magnetic fields of the underlying
cool disk penetrating into the corona leads to interesting aspects
for accretion in AGN. Our work suggests that the difference in
truncation radii derived for LLAGN and ellipticals with similar
accretion rates is due to the very different black hole masses,
$10^{6.5}M_\odot$ and $10^9 M_\odot$ respectively.
For the same mass flow rate and at the same distance from the black
hole (when measured in units of ${\dot M_{\rm {Edd}}}$ and  $r_{\rm
{S}}$) the disks in LLAGN are hot enough, but those in ellipticals
are cool and self-gravitating, no magnetic dynamo work. In LLAGN the magnetic
fields affect the coronal flow and shift the truncation to much smaller radii.
The truncation of disks in X-ray binaries and the spectral transitions
from hard to soft state (Meyer et al. 2000a) will also be affected
when disk dynamos occur.

\end{document}